\documentclass[final]{cimento}
\input psfig.tex
\def\lsim{\lower.5ex\hbox{$\; \buildrel < \over \sim \;$}}
\def\gsim{\lower.5ex\hbox{$\; \buildrel > \over \sim \;$}}
\title{Advective Accretion Flows: Ten Years Later}
\author{S.~K.~Chakrabarti\from{ins:x}}
\instlist{\inst{ins:x} S. N. Bose National Centre For Basic Sciences
JD Block, Salt Lake, Sector-III, Calcutta-700091, India,
email: chakraba@boson.bose.res.in}
\PACSes{\PACSit{04.70.-s}{Physics of black holes}
\PACSit{97.10.Gz}{Accretion and accretion disks}
\PACSit{98.38.Fs}{ets, outflows, and bipolar flows}
\PACSit{33.20.Rm}{X-ray spectra}}
\begin{document}

\maketitle

\begin{abstract}

Ten years have passed since the global solutions of advective accretion disks 
around black holes and neutron stars were first discovered. Since then 
they are enjoying support from observers almost on a daily basis, more so 
in recent days with the launching of very high resolution satellites. 
This review presents the development of the subject of advective accretion 
in last twenty five years leading to the global solutions and their applications.
It also shows that apart from the standard Keplerian disk features in 
most part of the accretion flow, future models must incorporate the essential 
features of the advective disks, such as the advection of energy and entropy
by the flow, centrifugal barrier supported boundary layer of a black hole,
steady and non-steady shocks, the bulk motion Comptonization of matter 
close to the black hole, outflows from the centrifugal barrier etc.
Since black holes are `black', methods of their 
identification must be indirect, and therefore, the solutions must be
known very accurately. On the horizon, matter moves supersonically,
but just before that it is subsonic due to centrifugal
pressure dominated boundary layer or CENBOL where much of the infall energy is
released and outflows are generated. In this review, we show that
advective flow models treat accretion and winds onto black holes and 
neutron stars in the same footing. Similarly treated are the steady and 
time-dependent behaviour of the boundary layers of neutron stars and the {\it black holes}!
Several observational results are presented  which 
support the predictions of this advective accretion/outflow  model. 
Signals from coalescing gravitating waves are affected by the 
sub-Keplerian flows as well.

\end{abstract}

\noindent{Il Nouvo Cimento, In press)

\section{Introduction}

Standard accretion disk models of Shakura \& Sunyaev~\cite{ref:ss73} 
and Novikov \& Thorne~\cite{ref:nt73} have been very useful in interpretation
of observations in binary systems and active galaxies (e.g. \cite{ref:p81, ref:st84, ref:f92}). 
The description of physical quantities in these models are expressed analytically and they could be used 
directly. However, pressure and advection terms were not treated correctly, 
since the disk is terminated at the marginally sable orbit (three Schwarzschild 
radii for a non-rotating black hole). A second problem arose, when it was 
found out  that the hot {\it Keplerian disks} of  Shapiro, 
Lightman \& Eardley~\cite{ref:sle76} are viscously and thermally unstable close to the inner regions. 
Paczy\'nski and his collaborators \cite{ref:pb80, ref:pm82}
attempted to include advection and pressure effects in the so-called transonic accretion
disks, although no systematic study of global solutions was made.
Global solutions of the so-called `thick accretion disks' were possible to
obtain only when the advection term is dropped (e.g.,~\cite{ref:pw80}). 
In these accretion disks, the flow is assumed to have 
practically constant angular momentum. Some exact solutions of {\it fully general 
relativistic} thick disks are discussed in Chakrabarti~\cite{ref:c85}.

Meanwhile, observations of Cyg X-1 in the early seventies appears to indicate 
that it emits X-rays in two states~\cite{ref:a72, ref:t72}
which lead to speculations that probably the inner optically thin region 
of the disk emits the hard X-rays and the outer optically thick flows 
emit soft X-rays~\cite{ref:tp75}. Ichimaru~\cite{ref:i77} correctly judging 
that the advection would be important close to a black hole, obtained a new 
optically thin solution (which in the self-similar limit are known 
as Advection Dominated Accretion Flows of Narayan \& Yi~\cite{ref:ny94}) which includes 
heating, cooling and advection. This was done using a Newtonian potential.
Ichimaru found that depending on the outer boundary condition, the solution could
go over to the optically thick or the optically thin branch. There was
no global solution, however. In the presence of a black hole, matter must 
be supersonic on the horizon, and, early attempts to find global solutions 
of viscous transonic flow (VTF) equations~\cite{ref:pm82, ref:m84} 
concentrated much on the nature of the inner sonic point of these
flows which is located around the marginally stable orbit.
Meanwhile, in the case of inviscid adiabatic flow, an example of global 
solution was provided~\cite{ref:f87}
where study of shocks similar to that in solar winds and galactic jets~\cite{ref:f85}
was made. In the so-called `slim-disk' model~\cite{ref:a88},
using {\it local solutions}, it was shown that the instabilities 
near the inner edge could be removed by the addition of the 
advection term. This was done in optically thick limit. 
First satisfactory global solution of the governing equations in the
optically thin or thick limit which include advection, viscosity, 
heating and cooling in the limit of isothermality condition was 
obtained by Chakrabarti~\cite{ref:ttaf90a, ref:mnr90b} where  disk
models of (single) temperature which become Keplerian far away were considered. 

In an earlier work~\cite{ref:c89}
complete classification of global solutions of an
{\it inviscid}, polytropic transonic flow (see, Fig. 4 of~\cite{ref:c89}) which showed 
that in some region of the parameter space, the flow will have multiple sonic
points~\cite{ref:lt80}. Within this 
region, there is a sub-class of solutions where Rankine-Hugoniot
shock conditions are satisfied and standing shock waves are formed due to the
centrifugal barrier. Four locations, namely, $x_{si},\  (i=1..4)$ were 
identified where these shocks could formally be possible, but it was
pointed out that only $x_{s2}$ and $x_{s3}$ were important for accretion 
on black holes since the flow has to be supersonic on a black hole horizon 
and $x_{s1}$ could also be important for a neutron star accretion 
while $x_{s4}$ was a purely formal shock location. In Chakrabarti~\cite{ref:ttaf90a, ref:mnr90b}
viscosity was also added and complete global solutions in isothermal VTFs
with and without shocks, were found. In the language of Shakura-Sunyaev~\cite{ref:ss73} 
viscosity parameter $\alpha$, it was shown that if
viscosity parameter is less than some critical value $\alpha_{cr}$,
the incoming flow may either have a continuous 
solution passing through the outer sonic point, or, it can have standing shock 
waves at $x_{s3}$ (following notations of Chakrabarti~\cite{ref:c89, ref:ttaf90a, ref:mnr90b}
if the flow allows such a solution in accretion.
For $\alpha >\alpha_{cr}$, a standing shock wave at $x_{s2}$
persisted, but the flow now had two continuous solutions --- one
passed through the inner sonic point, and the other through the
outer sonic point.  The one passing through the inner sonic point is
clearly slowly moving in most of the regions, and therefore is optically
thick. The one passing through the outer sonic point becomes 
optically thin at a large distance. A new branch of solutions
which connected these two pieces is the standing shock wave solution
which is partly optically thick (in the vicinity of the
post-shock region) and partly optically thin (pre-shock flow coming out of a
Keplerian disk). Later analytical and numerical works~\cite{ref:cm93, ref:cm95, ref:nh94}
showed that $x_{s3}$ is stable, and that for $\alpha>\alpha_{cr}$ the 
continuous solution passing through the inner sonic point is chosen. 
Most importantly, these solutions show that they could join (though
not quite smoothly, since viscosity etc. were chosen to be constant) 
with a Keplerian disk at some distances, depending on viscosity and 
angular momentum~\cite{ref:ttaf90a}.

Extensive numerical simulations of quasi-spherical, inviscid,
adiabatic accretion flows~\cite{ref:mlc94, ref:mrc96, ref:rcm97}
show that shocks form very close to the location
where vertically averaged model of adiabatic flows predict them~\cite{ref:c89}. The flow 
energy is conserved and the entropy generated at the shock is totally advected 
into the black hole allowing the flow to pass through the inner sonic point. 
It was also found, exactly as predicted~\cite{ref:c89}, that flows with positive energy 
and higher entropy form supersonic winds. In presence of viscosity also,
very little energy radiates away (e.g., Fig. 8 of ~\cite{ref:mnr90b}).
Having satisfied with the stability of these solutions~\cite{ref:cm93, ref:mlc94, ref:cm95,
ref:mrc96, ref:lmc98}, a unified scheme of accretion disks 
was proposed~\cite{ref:unam93, ref:cm95} which combines
the physics of formation of sub-Keplerian disks with and without shock
waves depending on viscosity parameters and angular momentum at the
inner edge. The solutions remained equally valid for both the black hole and neutron 
star accretions as long as appropriate inner boundary conditions are employed.

Subsequently, Chakrabarti~\cite{ref:gut96a} found that even when the 
isothermality condition is dropped, the flow topologies 
remained the same as in~\cite{ref:ttaf90a, ref:mnr90b}. Furthermore, it was shown that neutron star
boundary layer could be studied as the post-shock region of the incoming flow
itself (this was already pointed out~\cite{ref:c89}). The behaviour of the solution
with viscosity was found to be intriguing. At very low viscosity or very high viscosities
the flow does not have shocks, and sub-Keplerian, optically thin solutions emerge
out of the Keplerian disks, very similar to the Ichimaru~\cite{ref:i77} solution. However, when the
viscosity is intermediate, the solution passes through a standing shock for some region
of the parameter space when $\gamma < 1.5$ (a point completely ignored by workers who 
do not find these shocks thereby making the shocks `controversial').
For other regions there are no shocks where abrupt changes in thermodynamical 
properties take place but more gradual changes are seen in the centrofugal barrier.

Two dimensional solutions of advective disks were found in 1996
which showed the presence of funnels along the axis~\cite{ref:apj96b}.
Such funnels were non-existent in the Advection dominated solution 
(see, Chakrabarti~\cite{ref:oebhc98} for comparison of solutions.)
Similar conclusions, that the flow can't maintain its structure
along the pole has been reached by Paczy\'nski~\cite{ref:p98} and 
Park \& Ostriker~\cite{ref:po99} as well. They showed that there is no way matter can be kept hot
along the pole and matter must cool down. Exactly similar properties of the
outflows from CENBOL has been used to explain the low frequency QPO of the black hole
candiate GRS1915+105~\cite{ref:cm00, ref:yati99}.

Several other developments were taking place in the theory of advective flows
particularly in the MHD limit and when the flow is non-axisymmetric.
Chakrabarti~\cite{ref:wd90c} found all the global solutions of Weber-Devis type of
equations. In this case, the nature of the magnetic field is predefined to be
radial and azimuthal, but nevertheless the solutions indicated new types of
sonic points in both accretion and winds. Instead of three sonic points
(two X-types and one `O' type), one obtains five sonic points (three X-type: slow
magnetosonic, Alfv\'enic and fast magnetosonic and two `O' type). Shocks were 
also found analytically for the first time. Since then similar solutions
have been found for cold flows where pressure and gravity effects were ignored
(see, ~\cite{ref:t91} and subsequent works of this group).
However, till today, no new solution has been found other than those in C90c.

When the advective flow is non-axisymmetric, spiral shocks may form in some regions of the
parameter space~\cite{ref:s87, ref:spir90d}. The later solutions are height 
integrated. Numerical simulations indicated that these self-similar shock solutions 
are roughly correct~\cite{ref:sm92}.  Observationally, they may have been detected as 
well in M87. We shall discuss these observations later.

\section{Milestones in Advective Accretion Flow Solutions}

Table I summarizes roughly the way the 
subject of advective accretion flows was developed in last twenty-five years or so. 
All self-similar solutions and those which are special cases of these solutions
are excluded.  In Table II, we show the summary of the observational evidence for
the advective accretion flows.

\begin{table}
\caption{Milestones in Theoretical Works on advective accretion flow}
\label{tab:mils}
\begin{tabular}{lr}
\hline
Conclusions      & References     \\
\hline
Spectrum calculations using spherical flow & ~\cite{ref:shap73, ref:meso83, ref:babu89}\\
New optically thin branch of the solution &  ~\cite{ref:i77}\\
Multiplicity of sonic points  & ~\cite{ref:lt80}    \\
Systematic study of shocks in accretion  & ~\cite{ref:c89, ref:ttaf90a}  \\
Shocks in winds & ~\cite{ref:ttaf90a}\\
The accretion shock is stable & ~\cite{ref:cm93, ref:mrc96}\\
Classification of all inviscid accretion and wind solutions & ~\cite{ref:c89, ref:apj96b}     \\
Positive Bernoulli constant required for shocks and winds &~\cite{ref:c89}\\
Global solution of viscous advective flows & ~\cite{ref:ttaf90a, ref:mnr90b, ref:gut96a}\\
Change of solution topology in presence of viscosity & ~\cite{ref:mnr90b, ref:gut96a}\\
Fast- \& slow magnetosonic points in MHD flows & ~\cite{ref:wd90c}\\
Self-similar non-axisymmetric shocks &~\cite{ref:s87, ref:spir90d}\\
Unification of Keplerian and sub-Keplerian flows under one scheme & ~\cite{ref:unam93, ref:gut96a}\\
Post-shock flow has the similar properties as those of the hot ion tori & ~\cite{ref:unam93}\\
Gravitation waves emitted from binaries can be affected & ~\cite{ref:grav93}\\
Sub-Keplerian flows produce jets with more desirable properties & ~\cite{ref:cb92}\\
Vertically averaged flow model verified  by two-dimensional simulations & ~\cite{ref:mlc94}\\
Magnetic tension can cause disappearance of the inner disk & ~\cite{ref:tex94, ref:cd94, ref:dc94}\\
Necessity of a modified viscosity law in presence of advection & ~\cite{ref:cm95}\\
Keplerian/sub-Keplerian transition radius variation with viscosity & ~\cite{ref:ct95, ref:gut96a}\\
Explanation of the spectral states by sub-Keplerian/Keplerian flows & ~\cite{ref:ct95}\\
Explanation of hard tail in soft state by Bulk motion Comptonization & ~\cite{ref:ct95}\\
Two domensional solution of advective disks including shocks &~\cite{ref:apj96b}\\
Self-consistent  advective disk solutions in Kerr geometry & ~\cite{ref:apj96b, ref:mnr96}\\
Shocks may oscillate in a black hole accretion as in a White Dwarf accretion & ~\cite{ref:msc96}\\
Oscillating shocks can cause disappearance of the inner disk & ~\cite{ref:rcm97}\\
Shock oscillation as the explanation of Quasi-Periodic Oscillations & ~\cite{ref:msc96, ref:cm00}\\
ADAF solution are special cases of advective disk model & ~\cite{ref:gut96a, ref:cpr96, ref:lfy99}\\
Supersonic outflows are produced along funnel walls and at the transition radius 
&~\cite{ref:mlc94, ref:cetal96}\\
Ouflow production rate depends on compression ratio of the inflow & ~\cite{ref:ijpc98,
ref:oebhc98, ref:aac99}\\
Neutron disks can be produced around a black hole &~\cite{ref:skcbm99}\\
Nucleosynthesis can cause instabilities in advective disks &~\cite{ref:bmskc99}\\
Bound matter of negative energy can be accelerated to infinity &~\cite{ref:icskc00}\\
Global Inflow-Outflow Solutions (GIOS) &~\cite{ref:bang98, ref:oebhc98, ref:daschak}\\
Outflows can cause softening of spectra & ~\cite{ref:c98rap}\\
Outflow rate depends on the spectral states &~\cite{ref:yati99} \\
Outflows can cause quiescence states & ~\cite{ref:daschak}\\
\hline
\end{tabular}
\end{table}

\begin{table}
\caption{Observational evidence for advective accretion flow solutions}
\label{tab:mils}
\begin{tabular}{lr}
\hline
Conclusions  & References     \\
\hline
Hot radiation from accretion shock fits AGN spectra better & ~\cite{ref:cw92}\\
M87 Data ~\cite{ref:harm94} is best fitted with spiral shock solution & ~\cite{ref:m87c95}\\
Timing analysis matches with two-component advective disks & ~\cite{ref:cray96}\\
Bulk motion Comptonization explains hard tails in soft states & ~\cite{ref:bor99, ref:st99, ref:lt99}\\
Inner edge of the disk is around 10-20 Schwarzschild radii & ~\cite{ref:gil97, ref:dmat99, ref:n99}\\
Spectral evolution during novae outburst are explained & ~\cite{ref:c97}\\
Quasi-Periodic Oscillation (QPO) is due to oscillation of shocks & ~\cite{ref:cm00}\\
Duration of QPO is related to frequency of QPO & ~\cite{ref:cm00}\\
Outflows are produced within $100R_g$ of the disk & ~\cite{ref:jbl99}\\
\hline
\end{tabular}
\end{table}

\section{Schematic pictures of accretion and outflow processes}

\subsection{Building blocks of steady solutions}

With the advent of detailed and accurate observations which demand more accurate solutions 
for proper explanations, it has become necessary to revise models of accretion and outflows.
Figure 1 shows the classification of all possible steady, inviscid solutions of transonic flows
including schematic view of the disk around a black hole~\cite{ref:c89, ref:mnr96}.
This shows how an inviscid or weakly viscous flow look
like. Beside each solution a schematic picture is given of accretion and outflows. 
Inward pointing arrows indicate the accretion solution and outward pointing arrows indicate the
wind solutions. In (A) and (a), low energy and low angular momentum flow behaves like a 
spherical Bondi inflow and Parkar winds~~\cite{ref:enp60} 
well known in solar physics. In (B) and (b), a steady 
inflow or outflow solution does not have a shock but the inflow becomes hotter due to 
slowing down at the centrifugal barrier. A time dependent simulation shows 
that an oscillating shock forms~\cite{ref:rcm97}
which is accompanied by a non-steady outflow. In (C) and (c), the steady inflow 
solution has a standing shock and this has been verified by numerical simulations 
also~\cite{ref:cm93, ref:mlc94, ref:mrc96}. Positive energy outflow
is also found to be steady. In (D) and (d), the inflow passes through inner sonic point,
but the outflow now has a steady shock~\cite{ref:cm93}. 
In (E) and (e), neither the outflow nor the inflow has a 
shock in the steady solutions. But by analogy, it is expected that the
outflow would have an oscillatory shock. This has not been tested yet. In (F) and (f),
both the inflow and outflow have no shocks, and energetic steady solutions 
are present as both pass through the inner sonic point. In (G) and (H), incomplete solutions 
with  closed topology  are present. Thus no steady solution is possible. Simulations~\cite{ref:rcunp}
indicated that flow is highly unsteady in these cases
which are schematically shown in panels (g) and (h) respectively. Especially interesting 
is the solution of O*, whese disks could be `thick' not because of gas or radiation pressure,
but because of turbulent pressure.

\begin{figure}
\vbox{
\vskip 0.0cm
\hskip -0.0cm
\centerline{
\psfig{figure=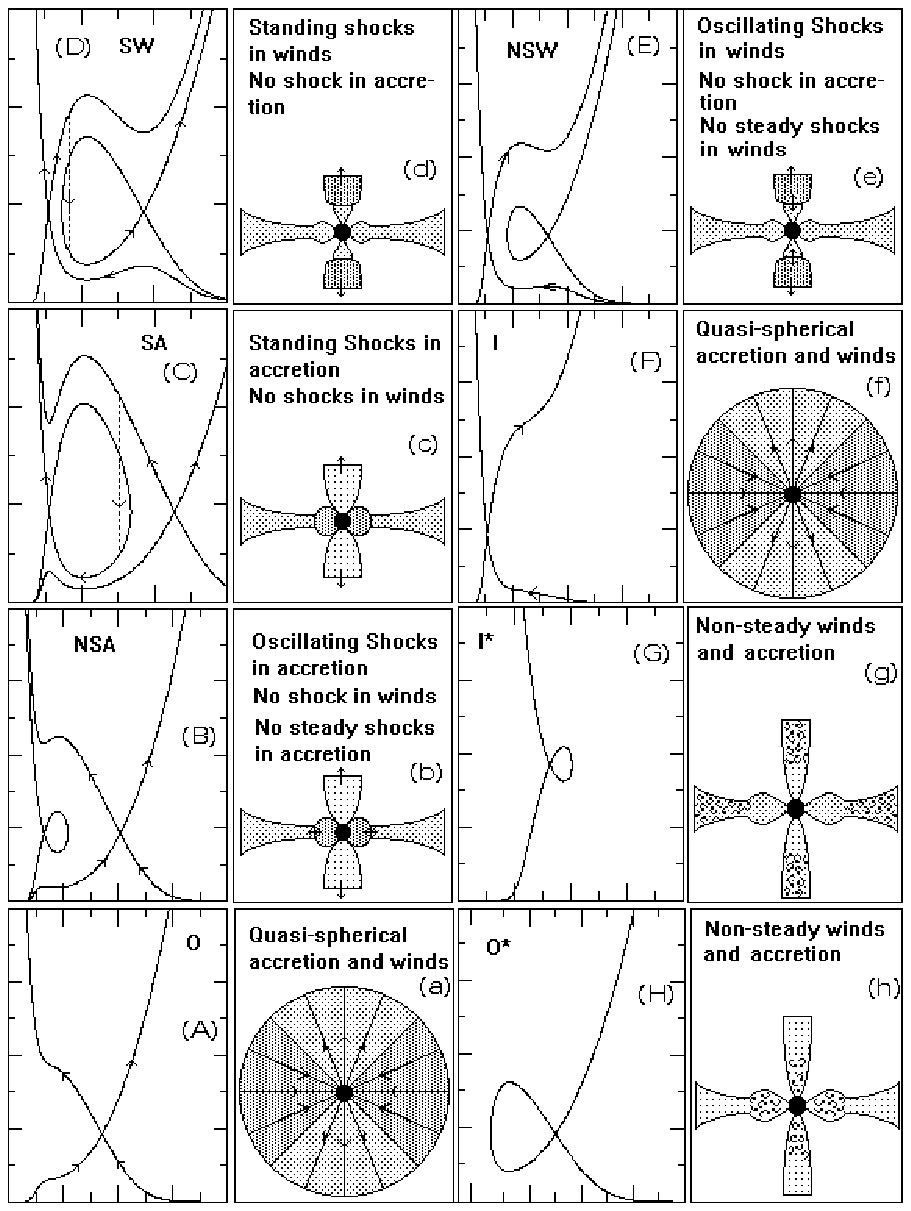,height=12truecm,width=10truecm}}}
\begin{verse}
\vspace{0.0cm}
\noindent {\small {\bf Fig. 1:} Nature of the solutions of inviscid advective disks and the corresponding 
flow behaviour. See text for details.}
\end{verse}
\end{figure}

\begin{figure}
\vbox{
\vskip 0.0cm
\hskip -0.0cm
\centerline{
\psfig{figure=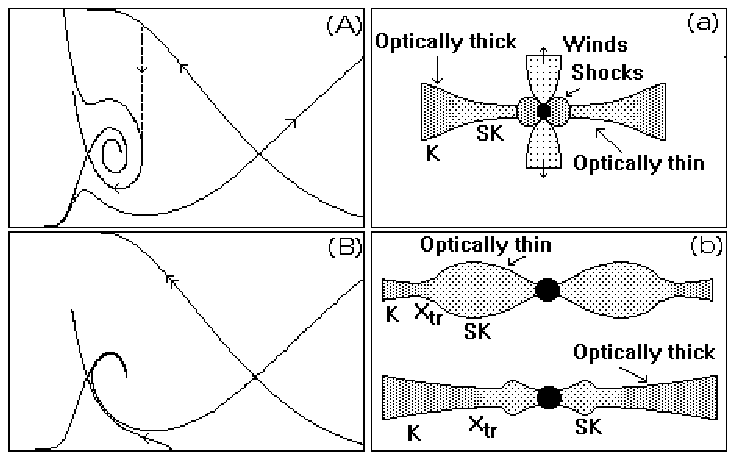,height=12truecm,width=12truecm}}}
\begin{verse}
\vspace{-6.0cm}
\noindent {\small {\bf Fig. 2:} Nature of the solutions of viscous advective disks 
and the corresponding flow behaviour: with shocks (A and a) and without shocks
(B abd b).}
\end{verse}
\end{figure}

In Figure 2, viscous transonic flow solutions are shown with constant viscosity parameter~\cite{ref:ttaf90a,
ref:mnr90b}.
When viscosity is low, the shock, albeit weaker, still forms (see, panel A) and 
the disk joins with a Keplerian disk far out. The flow at the transition (Keplerian to sub-Keplerian) 
radius need not be smooth (See, Fig. 3a for angular momentum distribution of the flow) and this may 
cause certain instabilities including oscillations and winds which carry away 
excess angular momentum~\cite{ref:cetal96, ref:korea96}. The nature of the disk corresponding to 
this is schematically shown in panel (a). In presence of higher viscosity, the closed topology of Fig. 1(B-C) 
opens up to accept matter straight from a hot flow provided the specific energy is positive.
Better choice is to have the flow away from an equatorial plane where specific energy is likely 
to be positive. In case the flow comes from a cooler Keplerian disk, energy is negative, and the topology I*
opens up to accept matter from it so that it may steadily enter the black hole through
the inner sonic point (Fig. 2B). The corresponding disk is schematically shown in Fig. 2b. In Fig. 3b, the
resulting angular momentum distribution is seen. If the viscosity is steadily reduced,
the transition radius recedes outwards. It is to be noted that the viscosity parameter $\alpha$
we are discussing here is only that is valid in the sub-Keplerian flow. It has no relation 
to the $\alpha$ parameter in the Keplerian  disk itself.

\begin{figure}
\vbox{
\vskip -5.0cm
\hskip 0.0cm
\centerline{
\psfig{figure=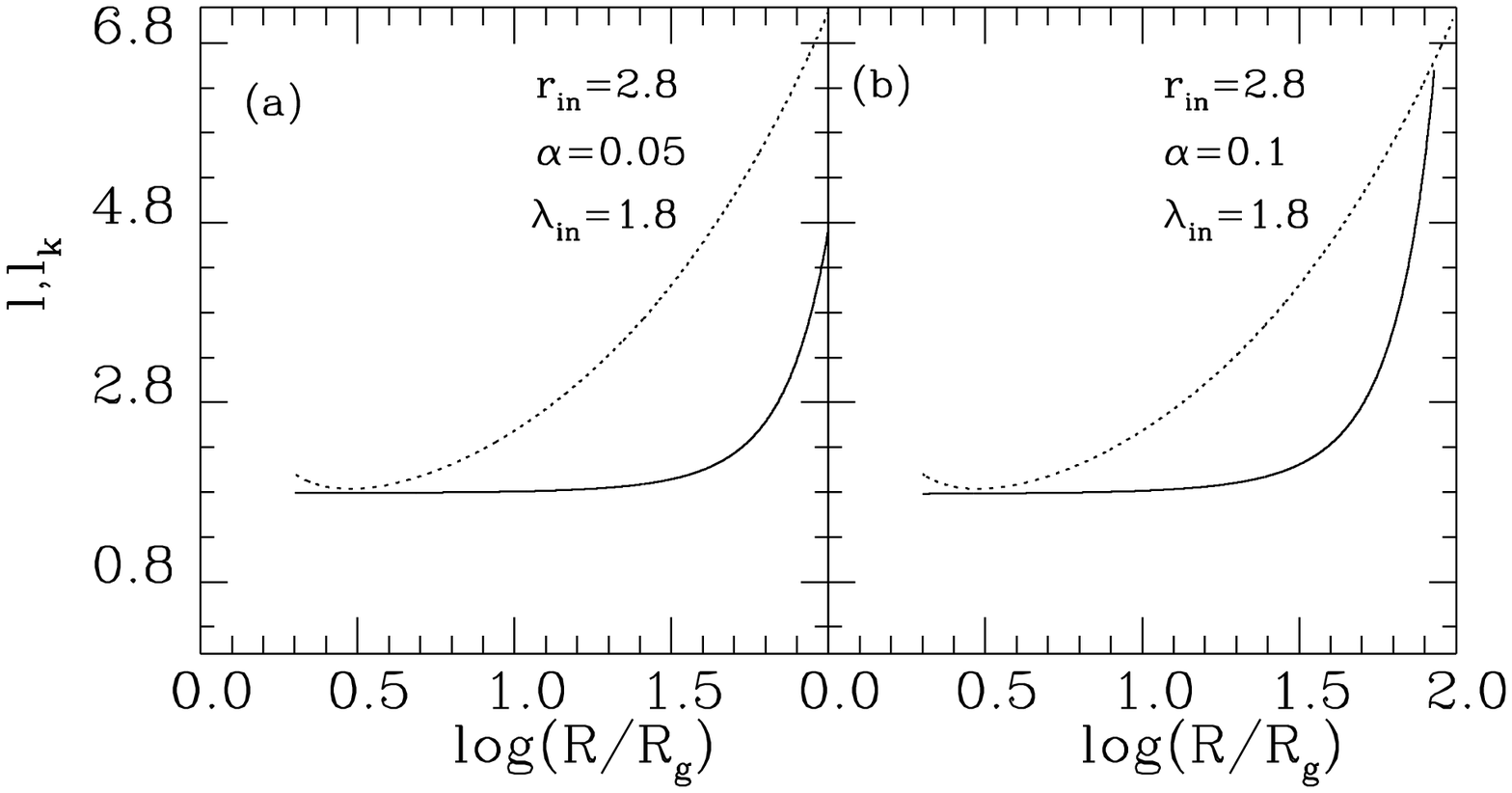,height=10truecm,width=12truecm}}}
\begin{verse}
\vspace{0.0cm}
\noindent {\small {\bf Fig. 3:} Nature of anguar momentum distribution (solid)
in an advective disk close to a black hole as a function of the logarithmic radial distance.
Two cases are same as (a) and (b) of Fig. 2. Dotted curves denote Keplerian distribution for
comparison.}
\end{verse}
\end{figure}

In realistic cases, neither the viscosity parameter is constant, nor the accretion rate
remains constant. More importantly, one needs to consider processes such as Coulomb 
coupling between ions and electrons, bremsstrahlung, Comptonizaion etc. Also important 
is the availability of the driving forces (such as thermal pressure, centrifugal, magnetic 
etc.) to form outflows. Depending on these factors, a realistic flow would be made up 
of combinations of these basic building blocks of the inflow and outflow.

One relavent matter in this regard is that ratio of the outflow rate to the inflow rate  $R_{\dot m}$
depends on the strength of the shock, when it forms ~\cite{ref:ijpc98, ref:aac99}.
One could argue that in soft states, when the accretion rate is 
high, the inner sub-Keplerian region is cooled down ~\cite{ref:ct95} and the shock is as good
as non-existent and there is no driving force to produce outflows ~\cite{ref:yati99}.
In the hard states, on the contrary, when the accretion rate is low, the stronger shocks
also causes very low outflow rates. In the absence of shocks, driving force 
is weak, and the outflow may be negligible (in the absence of strong magnetic fields).
For intermediate shock strength, the outflow rate is high and it could be cooled
down by soft photons from the pre-shock Keplerian disk. Once the flow is cooled down,
it would be super-sonic and fly away but the terminal velocity would be low.
This process may take place repeatedly ~\cite{ref:yati99, ref:cm00}
and cause low frequency variations of the light curve. In this so called `flare' state,
the outflow would be `blobby'. The variation of $R_{\dot m}$ with shock compression ratio 
is shown (solid) in Fig. 4 with a suggestive dependence of $R_{\dot m}$ shown on the 
spectral state of the flow. The dashed curve shows the variation of the actual outflow rate
when the accretion is assumed to vary as ${\dot M}_{in}=\frac{1}{R}$. This was done in 
keeping with the fact that in soft states ($R\sim 1$) the inflow rate is around ${\dot M}_{in} \sim {\dot M_{Edd}}$
and in very hard states ($R\sim 7$), ${\dot M}_{in} \sim 0.01 {\dot M_{Edd}}$.

\begin{figure}
\vbox{
\vskip -1.5cm
\hskip -0.0cm
\centerline{
\psfig{figure=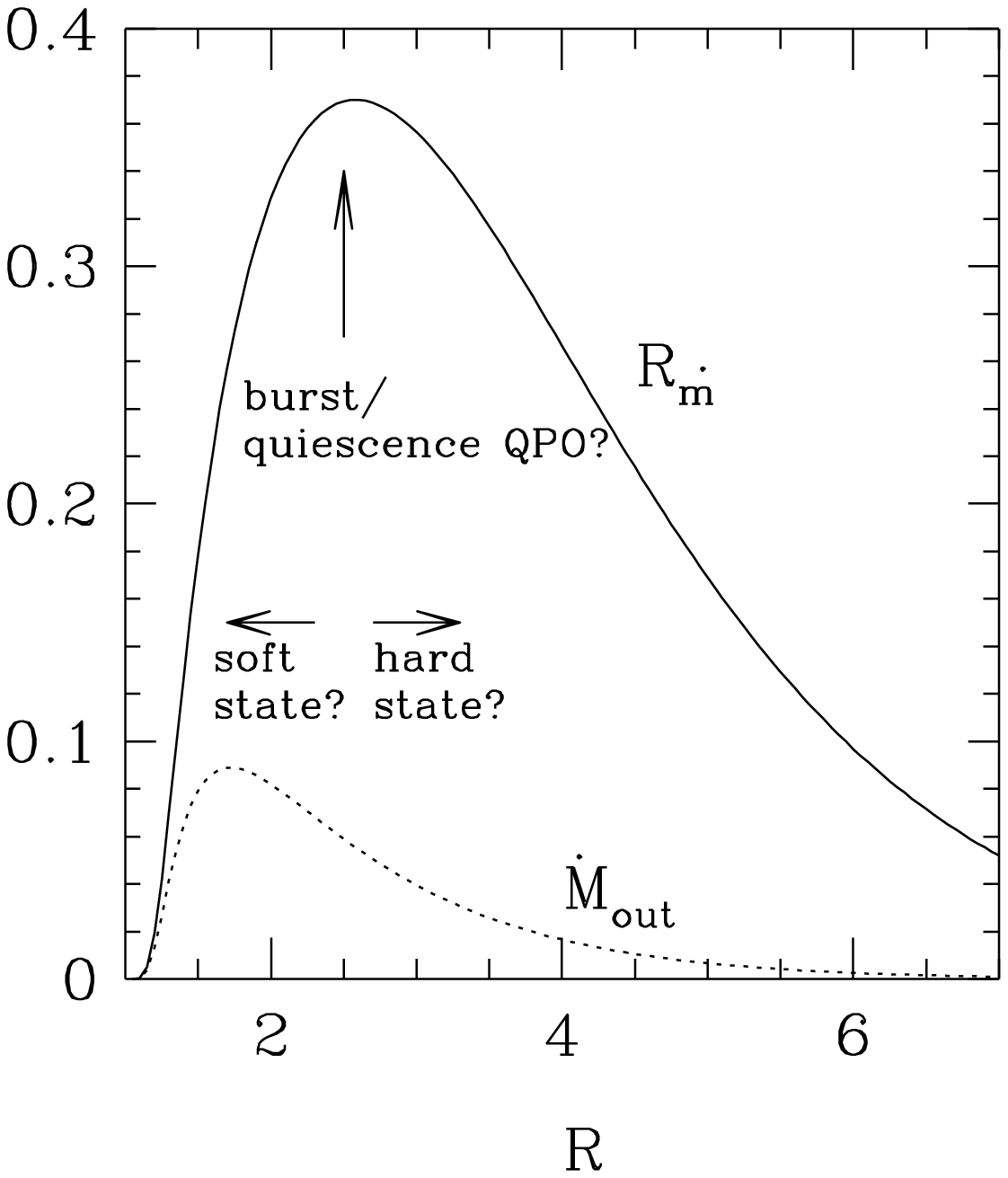,height=8truecm,width=10truecm}}}
\begin{verse}
\vspace{-1.0cm}
\noindent {\small {\bf Fig. 4:}
Variation of the ratio of the outflow rate to the inflow 
rate $R_{\dot m}$ with the compression ratio $R$ of the shock. It is believed 
that in soft states, $R\sim 1$ and thus the outflow rate is also very low. In
hard states, shocks are stronger and compression ratio is higher. 
Here, $R_{\dot m}$ is also low. For intermediate $R\sim 3$,
$R_{\dot m}$ is highest, outer sonic point of the flow $r_c$ is closest 
to the black hole. In this case, the outflow may be periodically 
cooled and it may therefore be blobby~\cite{ref:yati99, ref:cm00}. 
Dotted curve shows the actual outflow rate ${\dot M}_{out}$ when ${\dot M}_{in}$
varies as $1/R$ a plusible choice for an advective flow.}
\end{verse}
\end{figure}

\subsection{Building blocks of non-steady solutions}

Two major types of oscillations of the centrifugal pressure supported boundary layer is 
discussed in the literature, and it is possible that both are important. Figure 5 shows the
flow configurations at different times (measured in units of $2GM/c$). This happens when
the steady shock solution is absent even when the inner sonic point exists~\cite{ref:rcm97}.
In this case, the flow searches for a steady solution by 
first generating entropy through turbulence and then trying to pass through the inner sonic point 
non-existent in a steady flow. Figure 6(a-b) shows the nature of the variation of the entire disk
when bremsstrahlung cooling is added~\cite{ref:msc96}. This is
a prototypical case, when the steady shock solution exists but the cooling time 
in the post-shock/corona region roughly agrees with the infall time in this region.
An oscillation of similar kind is seen at the transition radius between Keplerian
and sub-Keplerian flows as some matter is removed as winds from this region~\cite{ref:cetal96, ref:korea96}. 
This is present even when no shocks are produced.
A third type of oscillation may be set in due to the quasi-periodic cooling of the   
outflow (provided it is high as is the case for average shock strengths $R\sim 2.5-3$ see, Fig. 4) due 
to enhanced interception of the soft-photons emitted from a Keplerian disk. 
It can be assumed that the quasi-periodic oscillation (QPO) in X-rays observed 
in black hole candidates is the result of one or more of 
these different types of oscillations since other types of oscillations, such as those due to
trapped oscillations~\cite{ref:khm88} or diskoseismology~\cite{ref:nvk93}
are incapable of modulating X-rays with large scale amplitude.

\begin{figure}
\vbox{
\vskip -1.5cm
\hskip -0.0cm
\centerline{
\psfig{figure=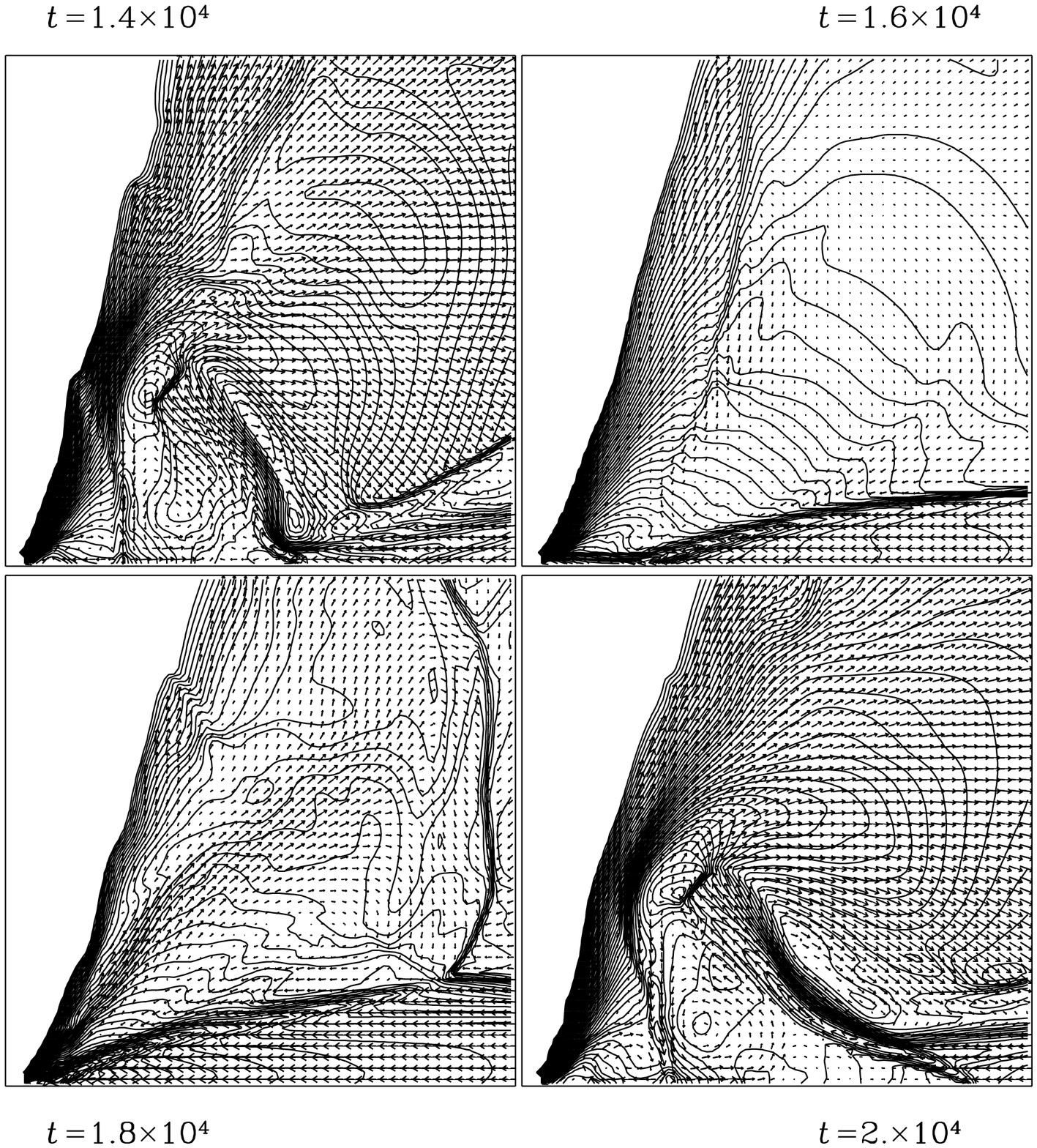,height=12truecm,width=12truecm}}}
\begin{verse}
\vspace{0.0cm}
\noindent{\small {\bf Fig. 5:} Oscillation of shocks and corresponding periodic changes in velocity 
field  when flow parameters are `wrong'. See~\cite{ref:rcm97} for details.}
\end{verse}
\end{figure}

\begin{figure}
\vbox{
\vskip 0.0cm
\hskip 0.0cm
\centerline{
\psfig{figure=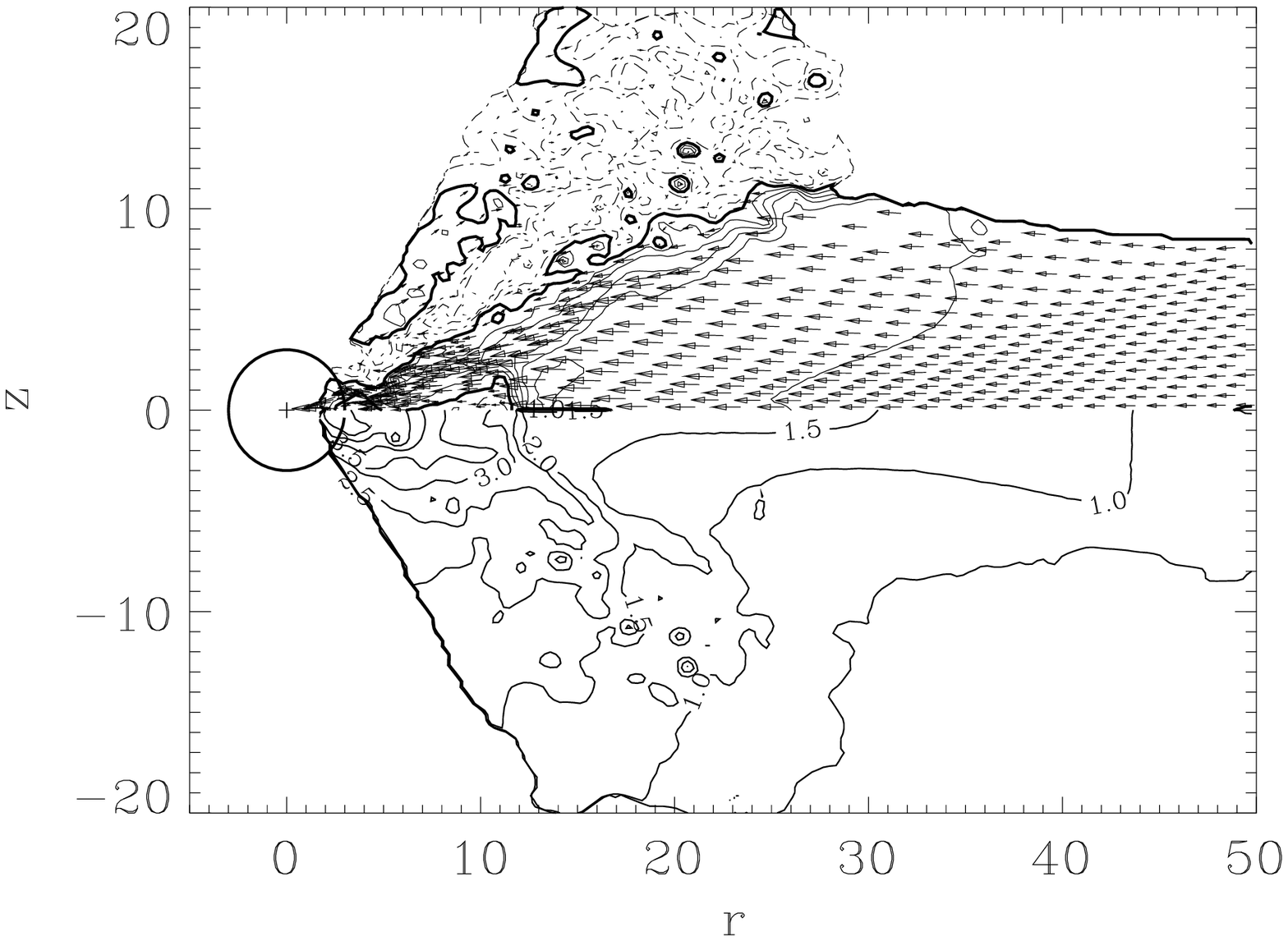,height=7truecm,width=8truecm}}}
\begin{verse}
\vspace{0.0cm}
\noindent{\small {\bf Fig. 6a:} Oscillation of shocks and corresponding periodic changes in velocity 
field when when flow parameters are such that the infall time matches with the cooling time. The corona
is prominent at this phase.
See~\cite{ref:msc96} for details.}
\end{verse}
\end{figure}

\begin{figure}
\vbox{
\vskip 0.0cm
\hskip 0.0cm
\centerline{
\psfig{figure=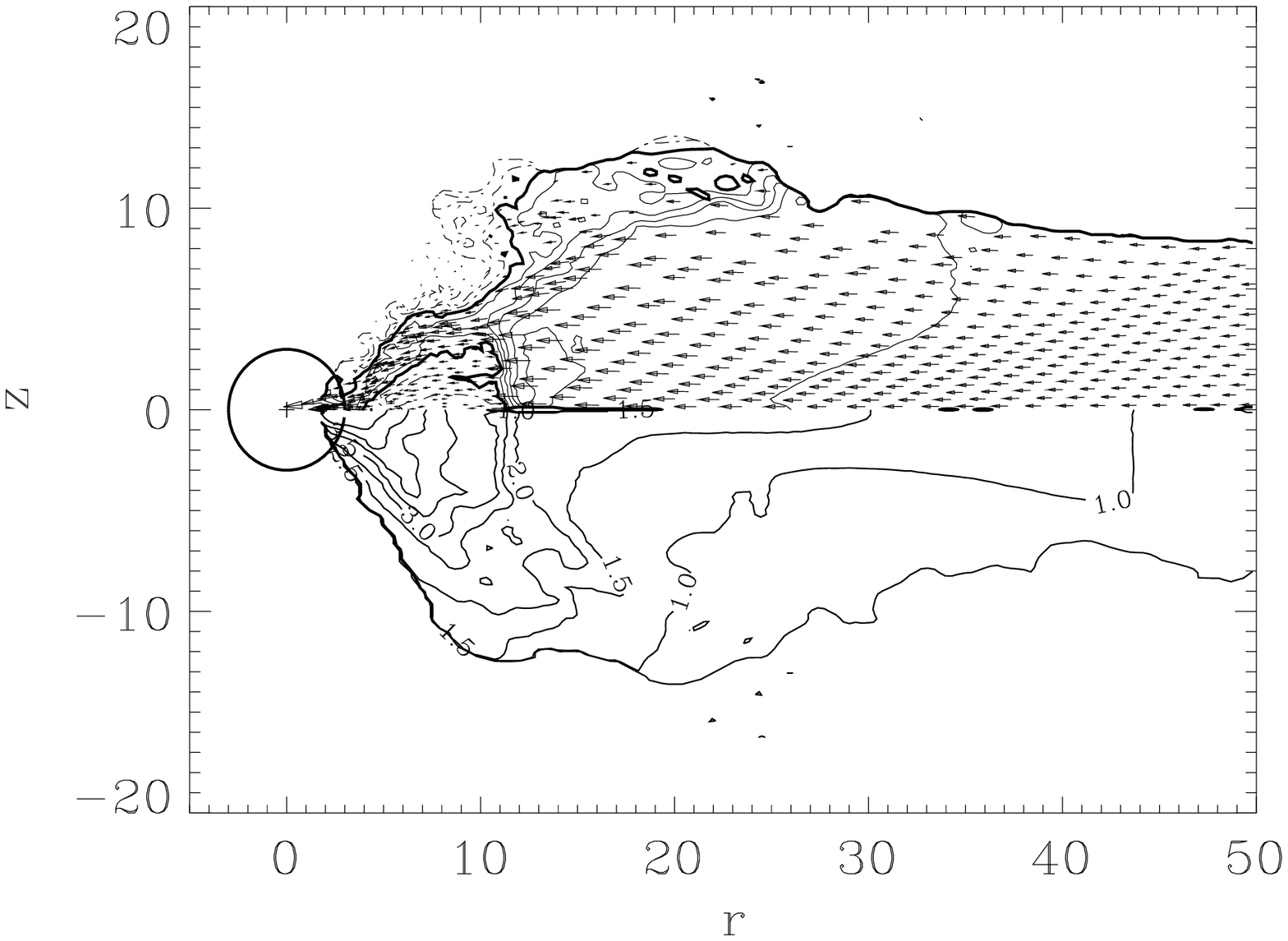,height=7truecm,width=8truecm}}}
\begin{verse}
\vspace{0.0cm}
\noindent{\small {\bf Fig. 6b:} Same as in Fig. 6a but after $T/2$ time when the corona is collapsed.
See~\cite{ref:msc96} for details.}
\end{verse}
\end{figure}

\subsection{Building blocks of magnetized accretion and winds}

All possible solution topologies of transonic flows in presence of magnetic field
are in Chakrabarti~\cite{ref:ttaf90a, ref:wd90c} and are not discussed here. The topologies would not change,
but the actual solution would be modified slightly when Grad-Shafranov equations are solved. 

\subsection{Combinations of building blocks}

One could combine results of Figs. 1-6 with to obtain realistic disk 
behaviour around a compact object.  Up until 5-6 years ago, it was 
believed that the hard radiation is the result of 
Comptonization of soft photons by `Compton Clouds' floating around 
the disk or by hot corona (Fig. 7). Chakrabarti and Titarchuk~\cite{ref:ct95}, for the first time
pointed out that the so-called `Compton Cloud' is the blotted-out, {\it sub-Keplerian} inner edge of the
Originally disk itself! (This is to be contrasted with the earlier models~\cite{ref:i77, ref:sle76}
which considered only the Keplerian disks.)
Thus, a new paradigm of accretion disks, based on actual solution
of advective flows, started. Today, this picture is universally adopted in most of the 
models of accretion flows. One can have one of the several types of flows
shown in Fig. 8. The so-called `hard' state is shown in Fig. 8(a-b)
and the so-called `soft' state is shown in Fig. 8c. The difference is that 
in the hard state, Keplerian flow rate could be very small while the sub-Keplerian
rate could be very high~\cite{ref:ct95}. In the soft state, it is the opposite. The de-segregation
of matter into these two types of rates are believed to be due to the fact that
the flow closer to the equatorial plane is likely to be more viscous and with larger
Shakura-Sunyaev parameter $\alpha$, and therefore is likely to be Keplerian. However,
flows away from the equatorial plane may have smaller $\alpha$ and therefore they deviate
from a Keplerian disk farther out~\cite{ref:gut96a}. Contribution to this sub-Keplerian flow may also have
come from wind accretion. This sub-Keplerian flow may form shocks at around
$10-20R_g$. In the soft state, shocks may be nominally present, but would be
cooler due to Comptonization, and would be as good as non-existent. The combined picture
in these two states was shown in Chakrabarti \& Titarchuk~\cite{ref:ct95}. As far as the
wind and outflow formation goes, the result strongly depends on the compression ratio ($R$)
of the flow at the shock as discussed in Chakrabarti~\cite{ref:oebhc98,ref:aac99} 
where an analytical expression
(Fig. 4) for the ratio $R_{\dot m}$ of the outflow (assumed isothermal) to inflow rates is provided.
When the compression ratio is unity, i.e., in the shock-free case (i.e., soft state),
the outflow rate is negligible. Thus, in soft states, no outflow is expected.
In the hard state, the shock is strong, and the outflow may be significant. In the
intermediate state outflow rate is highest. In Das \& Chakrabarti~\cite{ref:daschak}
it was shown that $R_{\dot m}$ should generally depend on inflow rates as well.
If a shock does not form and the accretion rate is low, the spectrum
will be harder and no quasi-periodic oscillations (QPO) would be seen~\cite{ref:c97}
in hard X-rays. This type of disks is shown schematically in Fig. 8a. 
If the shocks form, QPOs would be seen in hard X-rays, but softer X-rays,
arising out of a Keplerian disk will not be participating in the QPOs. This type of disk is schematically 
shown in Fig. 8b. When the viscosity and flow accretion rate is large, the sub-Keplerian region shrinks
as it is cooled down by thermal Comptonization. This is schematically shown in 
Fig. 8c. Only the bulk motion Comptonization 
will take place in flows at around $1-3R_g$ thereby producing hard tails in soft states~\cite{ref:ct95}.
Outflow also softens the spectra mimicking those from bulk motion Comptonization~\cite{ref:c98rap},
but as discussed above, it is unlikely that there would be strong outflows
in the soft state due to the lack of driving forces. 

\begin{figure}
\vbox{
\vskip 0.0cm
\hskip -0.0cm
\centerline{
\psfig{figure=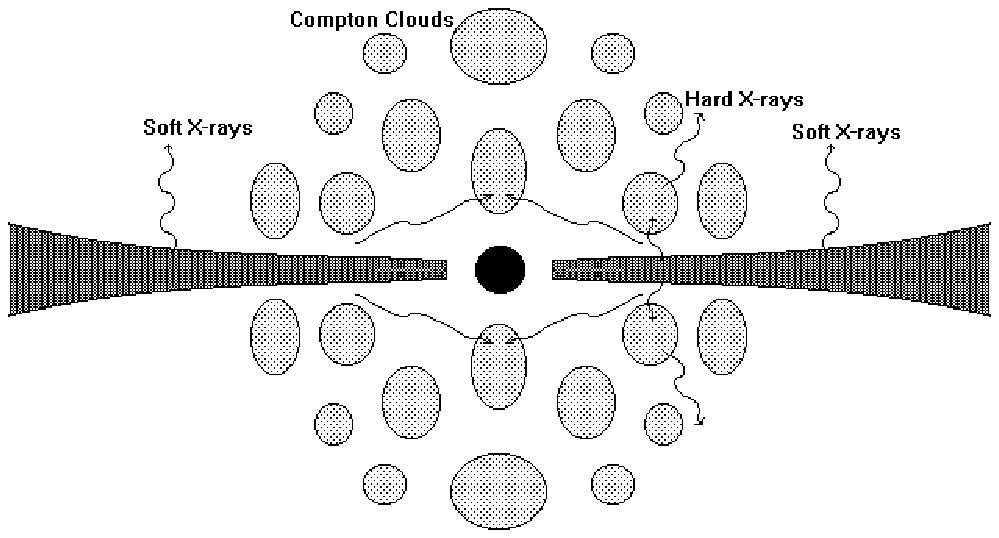,height=4truecm,width=6truecm}}}
\begin{verse}
\vspace{0.0cm}
\noindent{\small {\bf Fig. 7:} Cartoon diagram of a Keplerian disk with clouds of electrons
floating on both sides. This was the conception of black hole accretion flow before advective
disk solutions were applied.}
\end{verse}
\end{figure}

\subsection {Blobby Jets?}

We now discuss the conditions under which the jets could be blobby. Such jets were predicted purely from 
theoretical point of view, in the context of anti-correlation found in the X-ray and 
radio fluxes in the black hole candidate GRS 1915+105 (see, \S 4.7-4.8 of ~\cite{ref:cpr96}). Another easy way 
is to use magnetic tension. When magnetic field is dragged in by the accretion disks,
it could be sheared and amplified in the post-shock region till an equipartition is reached. A strong
field in a hot gas feels magnetic tension (`rubber-band effect') and is contracted catastrophically
evacuating the post-shock flow, i.e., the inner part of the accretion disk. Quoting ~\cite{ref:cpr96}:
`` Formation of coronae in an accretion disk is not {\it automatic}, i.e., presence of magnetic field
inside a disk does not automatically imply that a hot magnetic coronae would form. The formation requires
the ability of the accretion disks to anchor flux tubes inside the disk. This means that the disk
should have an internal structure akin to the solar interior and the entropy must increase
outwards. If the entropy condition is proper, the coronae would form, otherwise it would come out of the
disk as a whole without causing any random flare. In former case, there would be sporadic flaring 
events on the disk surface as in the case of the sun, whereas in the latter case, the collapse of fields
in the funnel would cause destruction of the
inner part of the disk and formation of blobby radio jets. Detailed observation of GRS1915+105 shows these
features~\cite{ref:mr94}. Since the inner part of the accretion disk could literally disappear by this magnetic
process, radio flares should accompany reduction of X-ray flux in this objects. Since the physical
process is generic, such processes could also be responsible for the formation of jets in active
galaxies and similar anti-correlation may be  expected, though time delay effects are to be incorporated
for a detailed modeling.'' In the absence of the magnetic field, periodic evacuation by
oscillating shocks can also take place if the flow parameters are `wrong' (see, ~\cite{ref:rcm97}
for details). This phenomenon may have been observed very convincingly in GRS1915+105.
In addition, the outflow may be periodically cooled. In Fig. 8d, we show disks in a possible flare-state
as observed in GRS 1915+105. Here shock strength is intermediate and the outflow is blobby.

\begin{figure}
\vbox{
\vskip 0.0cm
\hskip -0.0cm
\centerline{
\psfig{figure=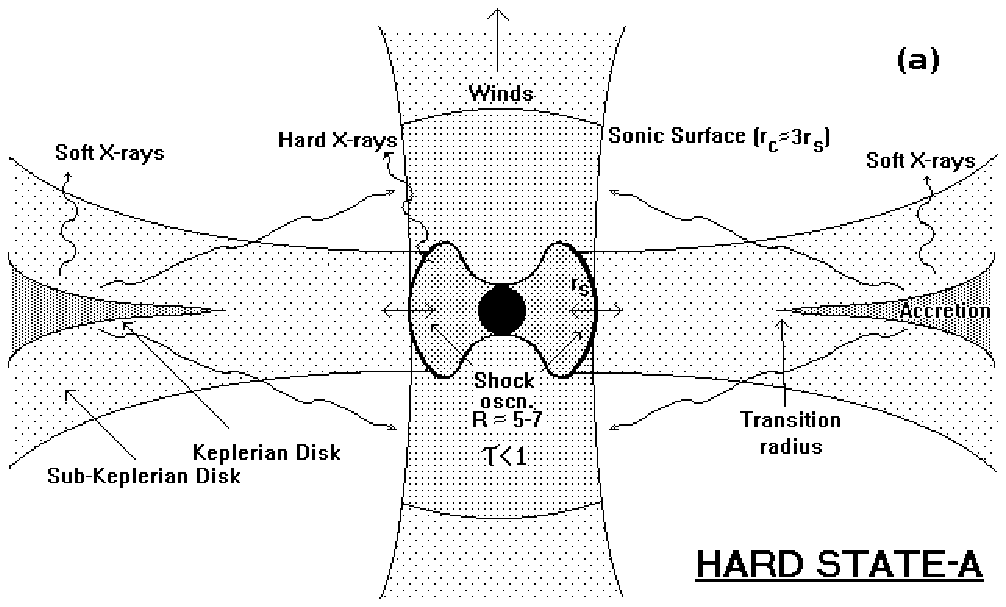,height=4truecm,width=6truecm}}}
\begin{verse}
\vspace{0.0cm}
\noindent{\small {\bf Fig. 8a:} Cartoon diagram of an advective disk when the building blocks
of Fig. 1(B-C) and Fig. 2(A-B) were used. A hard state is expected with or without QPOs. Wind
may be formed at a low rate (Fig. 4).}
\end{verse}
\end{figure}

\begin{figure}
\vbox{
\vskip +0.5cm
\hskip -0.0cm
\centerline{
\psfig{figure=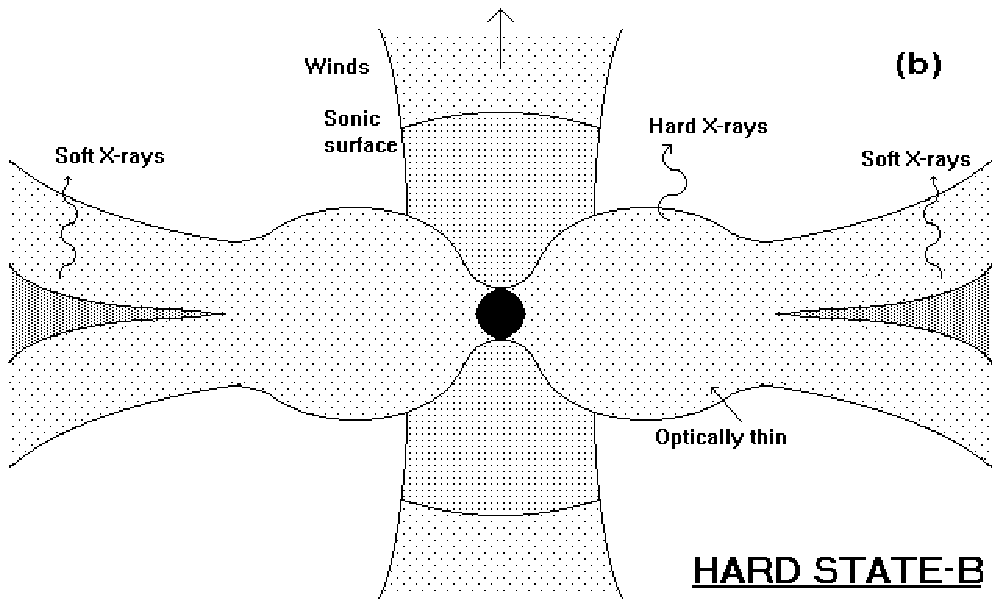,height=4truecm,width=6truecm}}}
\begin{verse}
\vspace{0.0cm}
\noindent{\small {\bf Fig. 8b:} Another possibility of hard states where shocks are not formed
in accretion. Building blocks of Fig. 1(A, D) and Fig. 2B were used. A hard state is expected without QPOs.
Winds may be formed at a low rate (Fig. 4).}
\end{verse}
\end{figure}

\begin{figure}
\vbox{
\vskip 0.5cm
\hskip -0.0cm
\centerline{
\psfig{figure=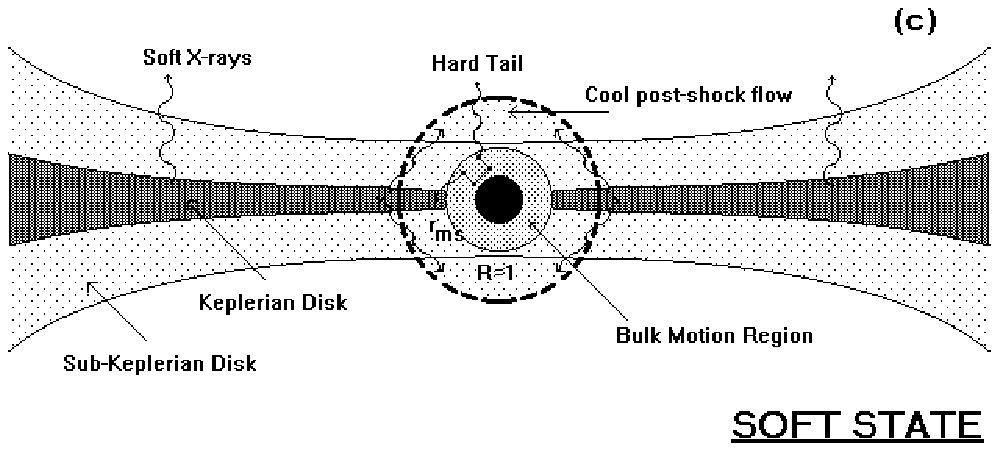,height=4truecm,width=6truecm}}}
\begin{verse}
\vspace{0.0cm}
\noindent{\small {\bf Fig. 8c:} Cartoon diagram of the advective disk with a 
soft state where shocks, if formed, are cooled down in accretion. 
Building blocks may be any of Fig. 1(A-F) and Fig. 2B were used. A 
soft state is expected without any QPO. N significant wind is possible (Fig. 4).}
\end{verse}
\end{figure}

\begin{figure}
\vbox{
\vskip 0.5cm
\hskip -0.0cm
\centerline{
\psfig{figure=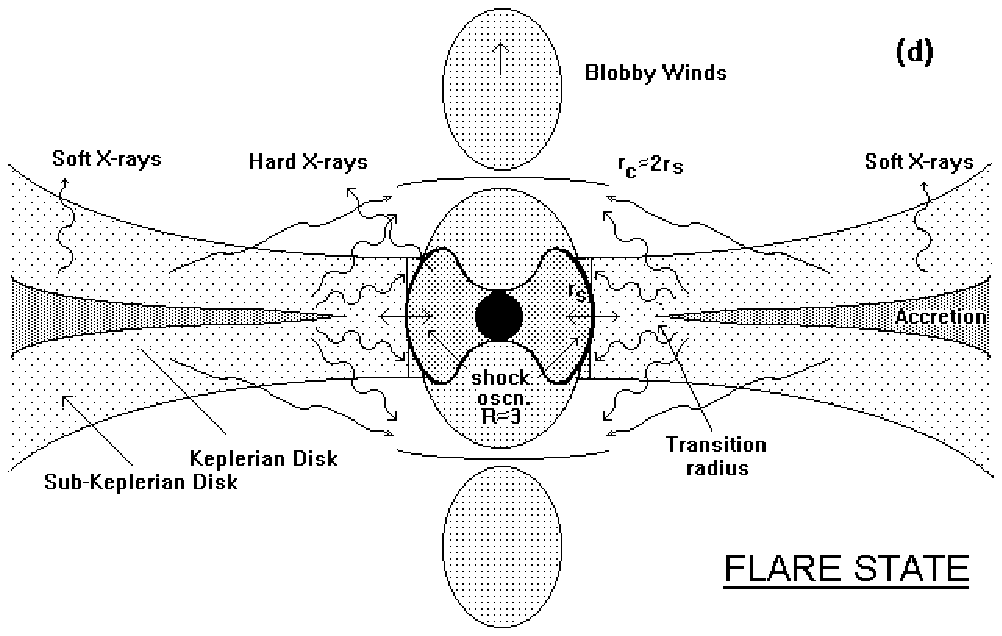,height=4truecm,width=6truecm}}}
\begin{verse}
\vspace{0.0cm}
\noindent{\small {\bf Fig. 8d:} Cartoon diagram of the advective disk in a flare state
where shocks of intermediate strength may form but since the 
outflow rate is high (Fig. 4), it may be periodically cooled to produce 
stange behaviour as in GRS1915+105. }
\end{verse}
\end{figure}

\section{Observations of Advective Accretion and Winds}

Several observations are now available which could be explained only with advective disks
since normal Keplerian disks appear to be insufficient.

\subsection{Sub-Keplerian motion on a large scale}
Since shocks are transitions from Supersonic to sub-sonic motion, and since supersonic
flows are sub-Keplerian (~\cite{ref:gut96a, ref:cpr96}), any presence of shocks
in a disk would indicate sub-Keplerian motions.
Sub-Keplerian flows rotate slower, and velocity predicted from Doppler shifted disk emission lines
would correspond to a higher Central mass. Chakrabarti~\cite{ref:m87c95} pointed out that the disk around M87
contained shocks and therefore the flow must be sub-Keplerian. As a result, the 
mass of the central object was found to be around $4\times 10^{9} M_\odot$ rather than $2 \times 10^9M_\odot$
as predicted by Harms et al. (see, ~\cite{ref:m87c95} for references) with 
calculations purely based on Keplerian motion.
The fact that shock-ionization causes the emission processes on M87 disk has been
stressed recently by several others~\cite{ref:dop97}. 

\subsection{Sub-Keplerian motion on a small scale}
It is usual practice to assume that the inner edge of a Keplerian disk extends till $3R_g$,
the marginally stable orbit. However, the advective disk 
models~\cite{ref:cpr96, ref:gut96a} show that the inner edge would be typically around
$\sim 10-20 R_g$ where the CENBOL should form. There are overwhelming evidence today that
this is indeed the case~\cite{ref:gil97, ref:n99, ref:dmat99}.
This sharply contrasts predictions of ADAF model which often
models the inner edge a distance of several tens of thousands (even at hundreds 
of thousands) of Schwarzschild radii!

\subsection{Power-law hard radiation in very high states}
Chakrabarti \& Titarchuk~\cite{ref:ct95} pointed out that when the accretion rate is relatively high,
the electrons in the sub-Keplerian region become cooler and this region becomes practically
indistinguishable from that of Keplerian disk. However, very close to the black hole horizon,
matter moves with almost velocity of light and deposits its bulk momentum onto the photons
thereby energizing these photons to very high energy forming a power-law. This power-law
is the hall-mark of all the known black holes~\cite{ref:bor99}. The success of this model
crucially hangs on the transonic flow solution  which utilizes the fact that the inner
boundary condition is independent of the history of incoming matter. Another possiblity of
producing hard spectra in soft states would be to use softening property of outflowing
winds~\cite{ref:c98rap} but since outflow should be weaker in soft states (Fig. 4), 
this procedure is not obvious.

\subsection{Quasi-Periodic Oscillations from black hole candidates}
X-rays from galactic black hole candidates often show persistent oscillations
which are quasi-periodic in nature.
Advective disk solutions do allow oscillations (Figs. 5-6), especially in the
X-ray emitting regions. The observation of these oscillations during the transition of states
is the triumph of the advective disk model~\cite{ref:rut99}. ADAF models, unlike advective disk solutions
described above do not allow shock formation~\cite{ref:n97} and therefore cannot address the 
issues of quasi-periodic oscillations at all.

\subsection{Outflows and their effects on the spectral properties}
As discussed in \S3.1 and \S3.2, there are outgoing transonic solutions which are counterparts of
transonic accretion flows onto black holes~\cite{ref:c89}. These outflows may
or may not have shocks, but they invariably have positive binding energy at a large distance.
ADAF solution recently came to the same conclusion (that Bernoulli
parameter should be positive~\cite{ref:qn97}). However, when the outflow is accelerated by external agency
such a conclusion is wrong as has been recently shown by Chattopadhyay and Chakrabarti~\cite{ref:icskc00}. 
They showed that even matter with negative initial binding energy could
be pushed to infinity (large distance) when radiation momentum is systematically
deposited on the flow! Recently, to overcome some of the shortcomings,
ADAF has also tried to produce some self-similar outflows (ADIOS~\cite{ref:bb99})
Being self-similar, jets are to come out all over the disk,
as opposed to regions close to the black holes as predicted by our advective disk
solutions. Recent high resolution observations of jets in M87 strongly suggest that
they are produced within a few tens of Schwarzschild radii of the horizon~\cite{ref:jbl99},
strongly rejecting ADAF and ADIOS models for the outflows,
which has no special length scale at these distances.
Given an inflow rate, one is now capable of computing the outflow rate
when the compression ratio at the shock surface is provided~\cite{ref:bang98, ref:aac99}.
This solution naturally predicts that the outflow must form at the CENBOL. The
Globally complete Inflow-Outflow Solutions (GIOS) were also found~\cite{ref:bang98}.
In presence of winds, the spectra is modified: outflows take away
matter and reduces the number of electrons from the post-shock (CENBOL) region
and thus they are cooled more easily by the same number of soft photons emitted
by the pre-shock Keplerian flow. This makes the spectra much softer~\cite{ref:ijpc98}
in presence of winds.

\subsection{Quiescence states of black holes}
Chakrabarti~\cite{ref:oebhc98} and Das \& Chakrabarti~\cite{ref:daschak}, 
pointed out that in some regions of the
parameter space, the outflow could be so high that it evacuates the disk and forms
what is known as the quiescence states of black holes. A well known example is the
starving black hole at Sgr A* at our galactic centre whose mass is
$\sim 2.6\times 10^6 M_\odot$ and the accretion rate is supposed to
be around $\sim 10^{-5} M_\odot$ yr$^{-1}$ which is much smaller compared to the
Eddington rate. Quiescence states are also seen in stellar mass black hole
candidates such as A0620-00 and V404 Cygni. Another way of producing these states
is to  use well known viscous instability in an accretion disk as used in models of dwarf-novae
outbursts~\cite{ref:ct95}.

\subsection{On and Off-states during QPOs in black holes}
The black hole candidates GRS1915+105 displays a variety of behaviour: usual high 
frequency QPOs are frequently interrupted by low frequency oscillations. 
While the QPO frequency can be explained by the shock oscillations, the switching of 
on and off states is explained by the duration in which extended corona becomes 
optically thick. It is possible that outflows are slowed down by this
process and the matter falls back to CENBOL, extending the duration of the `on' state
(which, if exists, is found to be comparable to the duration of the `off' state~\cite{ref:y99, ref:b97}).
The QPO frequency does evolve during this time scale and it is suggested that this is due to
the steady movement of the inner edge of the Keplerian accretion disk~\cite{ref:tru99}
or the steady movement of the shock itself in viscous time scale~\cite{ref:cm99}.
Generally speaking, shocks may show large scale oscillations~\cite{ref:msc96, ref:rcm97} but the wind plays a
vital role in intercepting larger number of soft photons, characteristics of soft state.
The correlation between the duration of the off-state and the QPO frequency has been
found to agree with the observations~\cite{ref:cm99}.

\subsection{Relationship of outflows and the black hole states}
Fig. 4 shows the possible relation between the outflow rates and the
spectral states of the black holes. Recently, from observational
point of view this picture has gained some support as well~\cite{ref:f99}.

\subsection{Effect of advective disks on gravitational wave}
It was pointed out~\cite{ref:grav93, ref:cprd96} that since advective disks
close to a black hole need not be Keplerian, it would affect the
gravitational wave properties of a coalescing binary. The angular
momentum loss to the disk by the companion in comparison to the angular momentum
loss to gravitation wave was found to be very significant~\cite{ref:cprd96}.
The variation of the signal is shown in ~\cite{ref:ijpc98}
One of the exciting predictions of this scenario is that since the spectrum of an
accretion disk contains a large number of informations (e.g. mass of the central black holes,
distance of the black hole, accretion rate, and viscosity parameter) a simultaneous
observation of the electromagnetic spectrum from the disk and the gravity wave spectrum
(which also must depend on those parameters, except possibly the distance)
should tighten the parameters very strongly. This issue is being investigated and will be
reported shortly.

Recently, ADAF model of  extremely low accretion rate was used to repeat
these computations, and not-surprisingly, no significant change
in gravitational signal was found~\cite{ref:n97}. If little matter is accreting,
it is as good as having no accretion disk at all. ADAF model ideally
valid for zero accretion rate systems and not found to be true in any of
the  realistic accreting systems. As discussed earlier,
evidences of the oscillating shocks, bulk motions etc. (which are absent in ADAF)
are abundant. Thus it is likely that the gravitational signal
{\it would be} affected exactly in a way computed by ~\cite{ref:cprd96}

\section{Concluding remarks}

When globally complete advective disk solutions were
first introduced ten years ago which included centrifugal barrier and 
standing shocks, it was considered to be of academic interest.
Today, with the advent of accurate observational
techniques, including those with high time resolutions,
there is enormous evidence that these are more realistic solutions
of the flow. There is hardly a single paper in the literature today which
does not use the advective disk solutions (sometimes not recognizing 
these point at all as in ADAF models.) As we have already noted, 
most of these improved observations cannot be explained at 
all using standard Keplerian disk or any simple variation
of it. We believe that all future models of the disks must include the
basic properties of this model including centrifugal barrier, standing shocks,
outflowing winds, sub-Keplerian inner region etc.

\acknowledgments
The author thanks DST for a partial support for the project `Analytical and numerical
studies of astrophysical flows around black holes and neutron stars' and the organizers
to partly supporting his trip to Rome and Pescara and covering the local hospitalities.

\end{document}